\documentclass[twocolumn,apl,amsmath,amssymb,showpacs,superscriptaddress]{revtex4-1}
\usepackage{epsf}      
\usepackage{graphicx}
\usepackage{color}
\usepackage{soul}
\usepackage{gensymb}
\usepackage{sidecap}
\usepackage{amsmath}
\usepackage{mathtools}

\begin{document}
\title{Unraveling diffusion kinetics of honeycomb structured Na$_2$Ni$_2$TeO$_6$ as a high-potential and stable electrode for sodium-ion batteries}
\author{Jayashree Pati}
\affiliation{Department of Physics, Indian Institute of Technology Delhi, Hauz Khas, New Delhi-110016, India}
\author{Hari Raj}
\affiliation{Department of Physics, Indian Institute of Technology Delhi, Hauz Khas, New Delhi-110016, India}
\author{Simranjot K. Sapra}
\affiliation{Department of Physics, Indian Institute of Technology Delhi, Hauz Khas, New Delhi-110016, India}
\author{Anita Dhaka}
\affiliation{Department of Physics, Sri Aurobindo College, University of Delhi, Malviya Nagar, New Delhi-110017, India}
\author{A. K. Bera}
\affiliation{Solid State Physics Division, Bhabha Atomic Research Centre, Mumbai 400085, India}
\affiliation{Homi Bhabha National Institute, Anushakti nagar, Mumbai, 400 094, India}
\author{S. M. Yusuf}
\affiliation{Solid State Physics Division, Bhabha Atomic Research Centre, Mumbai 400085, India}
\affiliation{Homi Bhabha National Institute, Anushakti nagar, Mumbai, 400 094, India}
\author{R. S. Dhaka}
\email{rsdhaka@physics.iitd.ac.in}
\affiliation{Department of Physics, Indian Institute of Technology Delhi, Hauz Khas, New Delhi-110016, India}
\date{\today}      

\begin{abstract}

In search of the potential cathode materials for sodium-ion batteries and to understand the diffusion kinetics, we report the detailed analysis of electrochemical investigation of honeycomb structured Na$_{2}$Ni$_{2}$TeO$_{6}$ material using cyclic voltammetry (CV), electrochemical impedance spectroscopy (EIS), galvanostatic charge-discharge (GCD) and galvanostatic intermittent titration technique (GITT). We found the discharge capacities of 82 and 77 mAhg$^{-1}$ at 0.05~C and 0.1~C current rates, respectively, and the mid-working potential of $\approx$3.9~V at 1~C and high capacity retention of 80\% after 500 cycles at 0.5~C as well as excellent rate capability. The analysis of CV data at different scan rates reveals the pseudo-capacitive mechanism of sodium-ion storage. Interestingly, the {\it in-situ} EIS measurements show a systematic change in the charge-transfer resistance at different charge/discharge stages as well as after different number of cycles. The diffusion coefficient extracted using CV, EIS and GITT lies mainly in the range of 10$^{-10}$ to 10$^{-12}$ cm$^{2}$s$^{-1}$ and the de-insertion/insertion of Na$^+$-ion concentration during electrochemical cycling is consistent with the ratio of Ni$^{3+}$/Ni$^{2+}$ valence state determined by photoemission study. Moreover, the post-cyclic results of retrieved active material show very stable structure and morphology even after various charge-discharge cycles. Our detailed electrochemical investigation and diffusion kinetics studies establish the material as a high working potential and long life electrode for sodium-ion batteries. 
\end{abstract}

\maketitle
\section{\noindent ~Introduction}

Since last three decades, rechargeable lithium-ion batteries (LIBs) have been widely used as an efficient power source for portable devices \cite{WhittinghamCR04, GoodenoughCM10, ManthiramNC20}. The high energy/power density of LIBs also steered their applications toward electric vehicles and grid scale energy storage systems. However, the ever-growing dependence on the portable devices in online world, increasing requirement of electric vehicles to keep the environment pollution free, as well as limited and non-uniform geographical distribution of lithium are the main reasons of high cost of LIBs \cite{ChoiACI12}. These factors raise a global call to find an alternative cost-effective solution for modern/common society. In this direction, sodium-ion batteries (SIB) having similar electro-chemistry attracted considerable attention due to the low cost as sodium is fifth most abundant element in the earth's crust, which may also solve the future geo-political issues related to lithium distribution \cite{HwangCSR17, DelmasAEM18, ChenMR18, LiEEC18}. In recent years a significant research is being carried out to develop electrode materials of SIBs and achieve the electrochemical performance (specially energy/power density) comparable to the LIB technology \cite{KimAEM12, PanEES13, LiEEC18, WuMT18}. However, there are many challenges in the SIBs for practical applications due to larger ionic radius of Na-ion (1.02 \AA ~for Na$^{+}$ vs. 0.76 \AA~for Li$^{+}$), which usually results in sluggish reaction kinetics and affect the energy storage capacity, rate capability, and cycling stability \cite{PalomaresEES12}. Therefore, it is imperative to develop suitable electrode materials with large interlayer spacing for efficient Na$^{+}$ storage and reversible intercalation/de-intercalation process \cite{HanEES15}. In fact, many oxide materials have been explored as electrodes for SIBs including layered and polyanionic structures having high theoretical capacity and high operating potential, respectively \cite{MaheshCI18, RakeshACSO19, MaheshEA20, Simran21}. In order to improve the electrochemical performance of SIBs for practical use in energy storage devices, one should focus on to achieve high specific capacity and high potential simultaneously.       

In this context, some layered honeycomb-ordered compounds Na$_{x}${\it M}$_{2}${\it M}$'$O$_{6}$ ($x=$ 2, 3; $M=$ Co, Ni, Zn, Mg, Fe; $M'=$ Sb, Te, Bi) have thrust into the limelight as potential cathode materials for sodium-ion batteries because of their significant sodium conductivity, excellent chemical stability as well as relatively high capacity and operating potential \cite{EvstigneevaCM11, YuanAM14, ZhengJES16, LiACS-AMI18, BhangeJMC17, WangJMCA19, YouAEM17, AguesseJPS16, WangACS17, WangNano18, GuptaJPS13, ChenCC20, GrundishCM20}. The honeycomb structure consists of alternative slabs of edge shared $M$O$_{6}$ and $M'$O$_{6}$ octahedra forming ($MM'$O$_{6}$)$^{x-}$- layers with intervening Na-layers. Moreover, the weak interlayer bonding creates vacancies in ($MM'$O$_{6}$)$^{2-}$ slabs, providing ease diffusion of Na-ions within the layers. Interestingly, the Na$_3$Ni$_2$SbO$_6$ cathode, having Ni$^{2+}$/Sb$^{5+}$ ordering, exhibits a discharge capacity of 110 mAh g$^{-1}$ at 1~C and around 90 mAh g$^{-1}$ at very high rate of 30~C with good retention and stability \cite{YuanAM14}. Further, the Zn doped Na$_3$Ni$_2$SbO$_6$ delivers the specific capacity ($\approx$115 mAh g$^{-1}$ when discharge at 0.1~C) as well as stability, and an average potential of around 3.3~V along a low polarisation as compared to the un-doped cathode \cite{AguesseJPS16}. The effect of Mg$^{2+}$ substitution at Ni site in Na$_3$Ni$_2$SbO$_6$ was found to suppress the P3--O1 phase transition and reduce volume expansion during sodium intercalation/de-intercalation process \cite{YouAEM17}. Also, the Mg$^{2+}$ being slightly larger than Ni$^{2+}$, the lattice parameters found to increase, which help in smoother sodium diffusion and improved kinetics resulting in better electrochemical performance at optimized Mg concentration \cite{YouAEM17}. Moreover, Yang {\it et al.} replaced one Ni with two sodium to increase its content, which shows the capacity of around 111 mAh g$^{-1}$ at 0.1~C and the working potential above 3.5~V due to inductive effect \cite{YangJPS17}. Also, the high ionic conductivity of Na$_2$Mg$_2$TeO$_6$ at room temperature makes it suitable for solid electrolyte in all solid-state batteries \cite{LiACS-AMI18}. A top-down synthesis approach was used to optimize the particle size/shape of Na$_3$Ni$_2$BiO$_6$ cathode for higher potential and better performance \cite{WangJMCA19}. Wang {\it et al.} studied the electrochemical performance of Na$_3$Ni$_{1.5}$$M_{0.5}$BiO$_6$ ($M=$ Ni, Cu, Mg, Zn) with the substitution of divalent elements at Ni site, and found that the Na$_3$Ni$_2$BiO$_6$ composition shows better stability as well as rate capability \cite{WangACS17}. This Ni based electrode exhibits a discharge capacity of 106 mAh g$^{-1}$ at 0.05~C, which was found to be very close to its theoretical capacity \cite{BhangeJMC17}. In another case, the O$^{'}$3 phase of Na$_3$Ni$_2$SbO$_6$ delivers the specific capacity of around 120 mAh g$^{-1}$ at 0.1~C along the high voltage and low polarization \cite{WangNano18}. A very recent review article provides detailed insights of physical/chemical properties of layered honeycomb compounds along the discussion about important factors for various applications \cite{KanyoloCSR21}.

Interestingly, the Na$_{2}$Ni$_{2}$TeO$_{6}$ compound having Ni$^{2+}$/Te$^{6+}$ ordering shows significantly high ionic conductivity and are considered as one of the high-voltage cathodes for SIBs \cite{GuptaJPS13}. In 2013, Gupta {\it et al.} first tested the electrochemical performance of Na$_{2}$Ni$_{2}$TeO$_{6}$ cathode and observe the capacity of around 90 mAh g$^{-1}$ at 0.05~C with high rate capability. Further, photoemission spectroscopy was used to confirm the valence state and amount of Na extraction corresponding to the observed voltage plateaus \cite{GuptaJPS13}. In fact, a lithium counterpart was reported to show higher voltage over 4~V due to the strong Te--O covalent bonding as well as inductive effect in more electronegative (TeO$_6$)$^{-6}$ anion \cite{GrundishCM19}. Recently, Bera {\it et al.} have reported increasing trend in the Na-ion conduction with temperature (0.03 S m$^{-1}$ at 423~K) and its pathways in the honeycomb structured Na$_{2}$Ni$_{2}$TeO$_{6}$, which make this material suitable as cathode for SIBs \cite{BeraJPCC20}. The sodium rich cathodes as well as effect of Co substitution at Ni site were also investigated in refs.~\cite{ChenCC20, GrundishCM20, YangJPS17}. However, detailed analysis of electrochemical investigation to understand the diffusion kinetics and long stability tests are still need to be explored for Na$_{2}$Ni$_{2}$TeO$_{6}$ cathode in SIBs.  

Therefore, in this paper we investigate the electrochemical performance and diffusion kinetics of honeycomb structured Na$_{2}$Ni$_{2}$TeO$_{6}$ (NNTO) material for sodium ion batteries using cyclic voltammogram (CV), electrochemical impedance spectroscopy (EIS), galvanostatic charge-discharge (GCD) and galvanostatic intermittent titration technique (GITT). Intriguingly, we found the discharge capacities of 82 and 77 mAhg$^{-1}$ at 0.05~C and 0.1~C current densities, respectively. The mid-working potential of around 3.9~V and high capacity retention of 80\% after 500 cycles at 0.5~C as well as excellent rate capability of Na$_{2}$Ni$_{2}$TeO$_{6}$ at various current rates demonstrate its potential as a stable cathode material for sodium-ion batteries. The cyclic voltammetry at different scan rates indicate both diffusion and capacitive controlled mechanism of charge storage. Also, using {\it in-situ} EIS we observe a systematic change in the charge-transfer resistance at different charge/discharge stages as well as after different number of cycles. More importantly, the diffusion kinetics study using CV, EIS as well as GITT analysis showed the value of diffusion coefficient mainly in the range of 10$^{-10}$ to 10$^{-12}$ cm$^{2}$ s$^{-1}$ depending on the experimental parameters like voltage. Understanding the change in structure and morphology of active material in electrode is crucial and we found Na$_{2}$Ni$_{2}$TeO$_{6}$  very stable using {\it ex-situ} XRD and FE-SEM measurements after $\ge$500 charge-discharge cycles. The notable outcome of our detailed analysis to unravel the diffusion kinetics demonstrate the potential of this material as a high potential and stable electrode for sodium-ion batteries. 

\section{\noindent ~Experimental}

The honeycomb Na$_{2}$Ni$_{2}$TeO$_{6}$ layered material was synthesized in single phase using solid state route. We use starting materials Na$_{2}$CO$_{3}$, NiO, and TeO$_{2}$ (all with 99.99\% purity) in stoichiometric ratio and heated the mixture at 950$^\circ$C in air for 72~h. The detailed procedure and other characterization are given in ref.~\cite{BeraJPCC20}. Before using the materials for battery applications, the phase purity of prepared powder was re-checked by X-ray diffractometer (Panalytical Xpert3) with CuK$\alpha$ radiation (1.5406~\AA) at the scan rate of 2$^\circ$/min in 2$\theta$ range of 10 to 90$^\circ$. The Rietveld refinement of XRD patterns were performed through Fullprof software for general crystal structure analysis. The morphology of prepared powder and electrodes after charge-discharge cycles was studied by field emission electron microscopy (FESEM). The gold coating was done on electrodes for conductivity before collecting FE-SEM data. The high resolution transmission electron microscopy (HR-TEM) measurements were conducted with Tecnai G2-20 system and analyzed using Image-J software. The ex-situ XRD pattern at different charged/discharged states were analyzed to elucidate the reversibility and stability of the material during electrochemical test. The Raman measurement was carried out with Renishaw invia confocal Raman microscope using 2400 lines/mm grating and 532~nm Argon laser with 1~mW power on the sample. The core-level spectra were recorded with an x-ray photoelectron spectrometer, PHI 5000 VersaProbe III using Al K$\alpha$ (h$\nu$ = 1486.6 eV) monochromatic x-ray source.

For the electrode preparation, we first prepare slurry by taking Na$_{2}$Ni$_{2}$TeO$_{6}$ as active material, carbon black as conducting source and polyvinylidene fluoride (PVDF) as binder in the ratio of 70:20:10 (by weight) and mixed in N-methyl-2 pyrrolidone (NMP) solvent. The slurry was casted (20~$\mu$m thick) on Al foil using doctor blade method and dried first in air for 12~h, then in vacuum at 120$^\circ$C for 8~h. The active material weight was taken around 2~mg cm$^{-2}$ in each cathode. We use 1M NaPF$_6$ electrolyte dissolved in EC and DEC (50:50 by volume), sodium metal foil as a reference electrode, glass fibre (GB100R) separator and 12~mm diameter of active electrode to fabricate CR2032 type coin cells inside the Argon gas-filled glove box (UniLab Pro SP from MBraun, Germany) having O$_{2}$/H$_{2}$O level $\leq$0.1~ppm. The CV at different scan rates, EIS, GCD at various current rates  and GITT measurements were performed using a Biologic VMP-3 system. A frequency range between 10 mHZ to 100 kHz is used for EIS measurement with a maximum voltage of 10~mV. For long cycling, a Neware battery cycler (BTS400) was used to collect GCD in the voltage range of 3.0--4.45~V. All the measurements were done at room temperature. The galvanostatic intermittent titration technique (GITT) measurements were performed with a current density of 0.1~C, where 5 min of current pulse was applied to the electrode followed by an open circuit relaxation of 30 min.

\section{\noindent ~Results and discussion}

\subsection{\noindent ~Structural and morphological characterization:}

In Fig.~1(a), we show the x-ray diffraction (XRD) pattern of prepared Na$_{2}$Ni$_{2}$TeO$_{6}$ material with the Rietveld refinement profiles and Bragg's positions. The refinement confirms the hexagonal layered P2-type crystal structure (space group: {\it P6$_3$/mcm}) having $a=b=$ 5.202~\AA, $c=$11.138~\AA~and $V=$ 261.02~\AA$^3$, other details and parameters are reported in ref.~\cite{BeraJPCC20}. In the inset of Fig.~1(b) we present the hexagonal crystal structure of Na$_{2}$Ni$_{2}$TeO$_{6}$ using VESTA software. The Raman spectrum between 100 to 800 cm$^{-1}$ wavenumbers is shown in Fig.~1(b), which shows the symmetric stretching modes of TeO$_{6}$ octahedra at 677 and 601 cm$^{-1}$. The bending modes of TeO$_{6}$ octahedra are centered at 481, 507, 651 and 665 cm$^{-1}$. The modes in the far-infrared region, i.e., 263 and 150 cm$^{-1}$ are due to streching of Na--O bond. The Raman shift of 560, 593 and 640 cm$^{-1}$ count the stretching vibrational modes of O--Ni--O bonds in NiO$_{6}$ octahedra \cite{KumarDT13}. Further, we checked the morphology and particles size of prepared powder sample by FE-SEM, as depicted in Fig.~1(c). 
\begin{figure}
\includegraphics[width=3.4in]{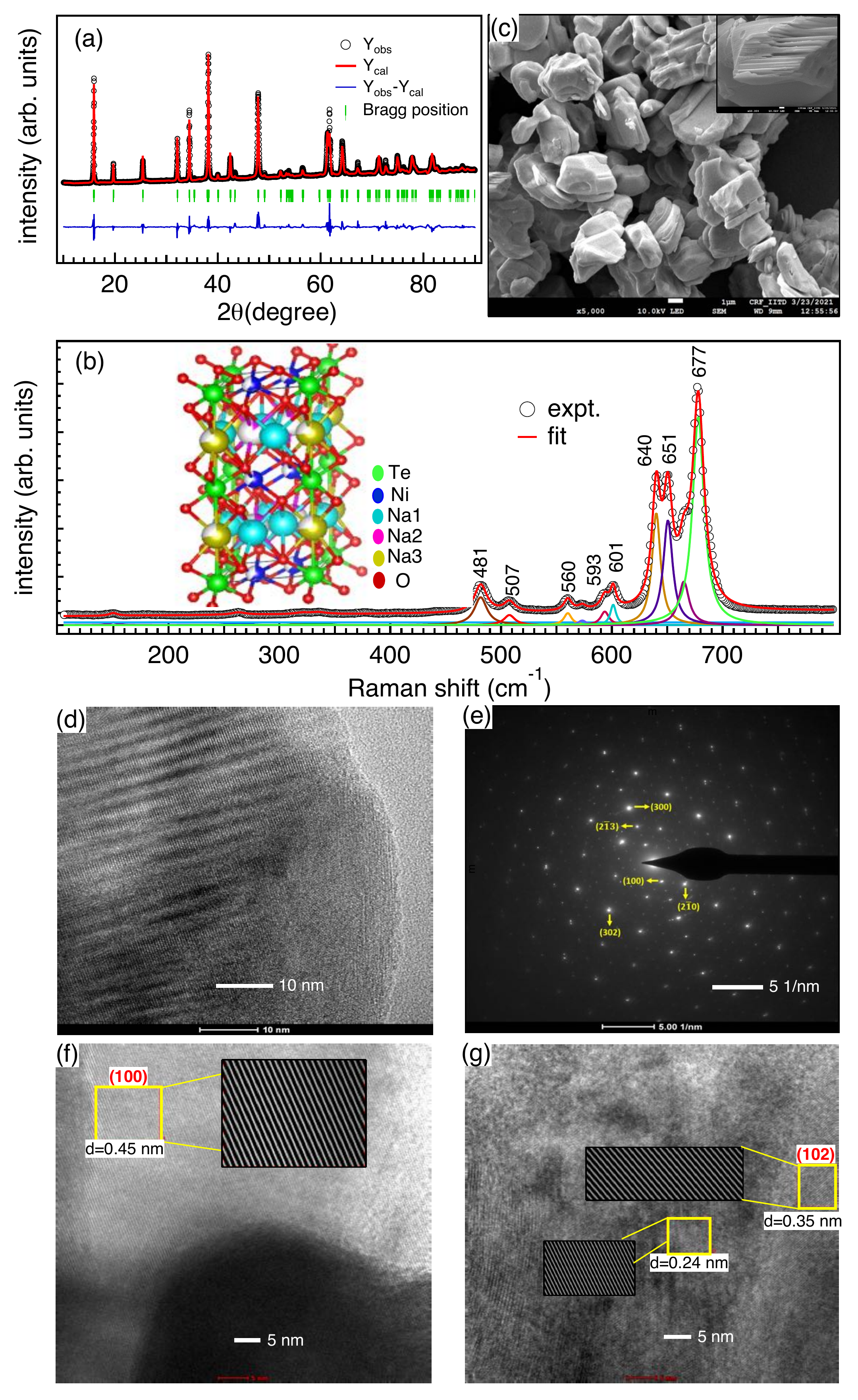}
\caption {Characterization of as-prepared Na$_{2}$Ni$_{2}$TeO$_{6}$ powder at room temperature. (a) The XRD pattern (black circles), the Rietveld refinement (red) and residual of fitting (blue) and Bragg's position (vertical green bars). (b) The Raman spectrum observed (black circles) and fitted (red line) along with the de-convoluted peaks corresponding to different modes. Inset in (b) shows the crystal structure where the Te, Ni and O atoms are represented by green, blue and red colors. The Na atoms at three crystallographically different sites are presented with cyan (Na1), pink (Na2) and yellow (Na3) color balls. (c) The FE-SEM image of same powder and the magnified view of a single particle in inset. (d) The HR-TEM image showing the layered morphology, (e) the SEAD pattern, (f) and (g) the HR-TEM images showing the interlayer spacing with corresponding indexed planes.} 
\label{DP_O_K}
\end{figure}
We observe uniform distribution of uneven shaped particles of 3--5~$\mu$m size, which is expected for samples prepared by solid-state reaction. Interestingly, the high magnification image in the inset of Fig.~1(c) reveals the layered structure of the material. The layered morphology of the materials would be very useful for electrolyte penetration, which also helps easy migration of Na-ions during electrochemical reaction. Moreover, the HR-TEM image in Fig.~1(d) shows the ridges in the electrode material and the SAED patterns with bright diffraction spots is shown Fig.~1(e), which represents the respective planes and well crystalline behaviour of Na$_{2}$Ni$_{2}$TeO$_{6}$ material. The HR-TEM images in Figs.~1(f) and (g) depict clear two-dimensional fringes with interlayer distance of 0.24, 0.35 and 0.45~nm between different crystal planes [corresponding to (112/104), (102), (100), respectively] of Na$_{2}$Ni$_{2}$TeO$_{6}$, which are greater than that of the ionic radius of Na$^{+}$ active ions and therefore expected to provide smooth intercalation/de-intercalation process.      

\subsection{\noindent ~Cyclic voltammetry (CV) and galvanostatic charge-discharge (GCD) investigation:}

In order to understand the sodium insertion and extraction mechanism we first perform CV measurements in the half-cell configuration against Na/Na$^{+}$ in the voltage window of 2.0--4.5~V. Fig.~2(a) shows three cycles of CV at a scan rate of 0.05 mV s$^{-1}$, which indicate multiple redox peaks at different voltages above 3~V \cite{ChenCC20}. The main oxidation peaks in the first cycle at around 3.78, 3.96, 4.3~V and the corresponding reduction peaks at around 3.52, 3.66, 3.94 are observed due to the multi-electron redox reaction associated with Ni$^{2+}$ to Ni$^{3+}$ couple \cite{YangAMI18, GrundishCM20}. Grundish {\it et al.} found that the electrochemical reactivity in Na$_{3}$Ni$_{1.5}$TeO$_{6}$ cathode is mainly due to the Ni$^{2+/3+}$ redox couple as well as small contribution from Ni$^{3+/4+}$ \cite{GrundishCM20}. 
The splitting in some of the oxidation/reduction peaks observed in the CV curves manifests the presence of different sodium ion disordering in the layered structure \cite{YangAMI18, WangSA18, MaseseMAT21}. The appearance of the minor redox peaks can be attributed to the phase transformations during Na$^+$ (de)intercalation reactions \cite{MaseseNC18}. In the honeycomb layered oxides, the multiple phase transitions have been reported in the voltage profile due to the Na ordering and transition metal gliding \cite{KanyoloCSR21, WangSA18}. The electrochemical Na insertion/extraction leads to the gliding of the honeycomb slabs, also called inter-slab gliding, which brings changes in the crystal structure upon cycling as Na atoms rearrange their occupying positions to prevent the structure collapse. These phase transitions lead to the appearance of the staircase like voltage profiles and is often described as the Devil's staircase \cite{KanyoloCSR21}. It is interesting to note that all the major anodic and cathodic peaks observed in CV are consistent with the voltage plateaus present in the galvanostatic charge discharge (GCD) profiles measured at 0.05~C current rate in the voltage range of 3--4.45~V, as shown in Fig.~2(b) for 10 cycles, which are as expected for materials having P2-type framework. We observe high reversibility in the GCD characteristics (except first charge) of the Na$_{2}$Ni$_{2}$TeO$_{6}$ cathode material and it shows a maximum  discharge capacity of 82 mAh g$^{-1}$, which retained around 77 mAh g$^{-1}$ upto 10th cycle. The multi plateau voltage vs time profile for 0.05~C current rate is depicted in Fig.~2(c). This stair case like feature of the profile signifies electrochemically-driven phase transition and multiple plateau also suggests a complicated phase transition \cite{ChenCC20, SuESM16} in Na$_{2}$Ni$_{2}$TeO$_{6}$ cathode at low current rate, as also clearly visible in Fig.~2(c). 

\begin{figure}  
\includegraphics[width=3.35in]{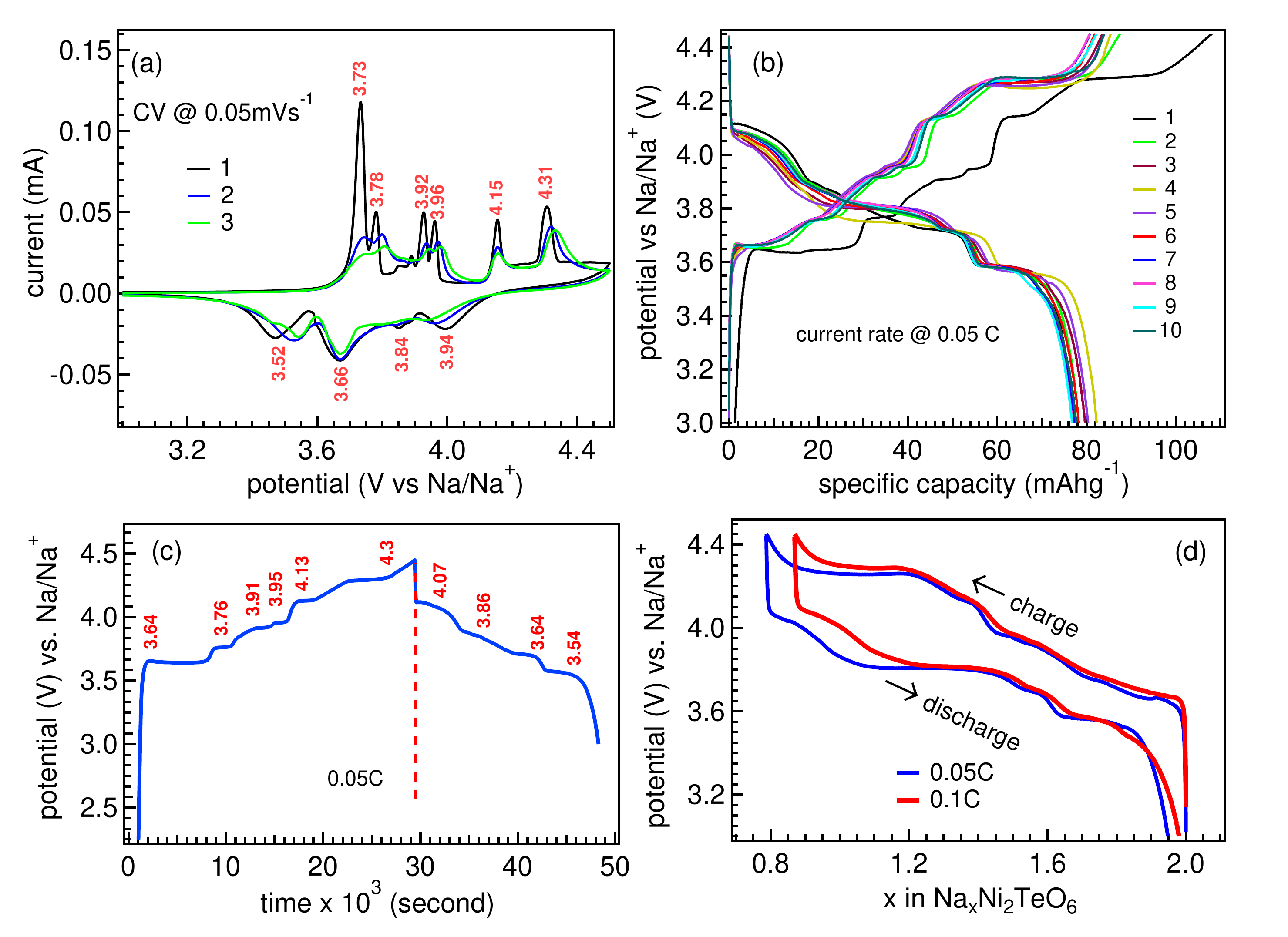}
\caption {The electrochemical characterisations of Na$_{2}$Ni$_{2}$TeO$_{6}$ cathode. (a) The cyclic voltammetry at 0.05~mVs$^{-1}$ for 3 cycles depicting the redox peaks in a voltage window between 2--4.5~V, (b) the galvanostatic charge-dicharge profiles at 0.05~C for first 10 cycles, (c) the voltage versus time curve showing multiple plateaus for phase transitions at particular voltages, (d) the voltage versus composition curve during 4$^{th}$ cycle for both charging and discharging at 0.05 and 0.1~C current rates.} 
\label{DP_O_K}
\end{figure}

\begin{figure*}  
\includegraphics[width=7.1in]{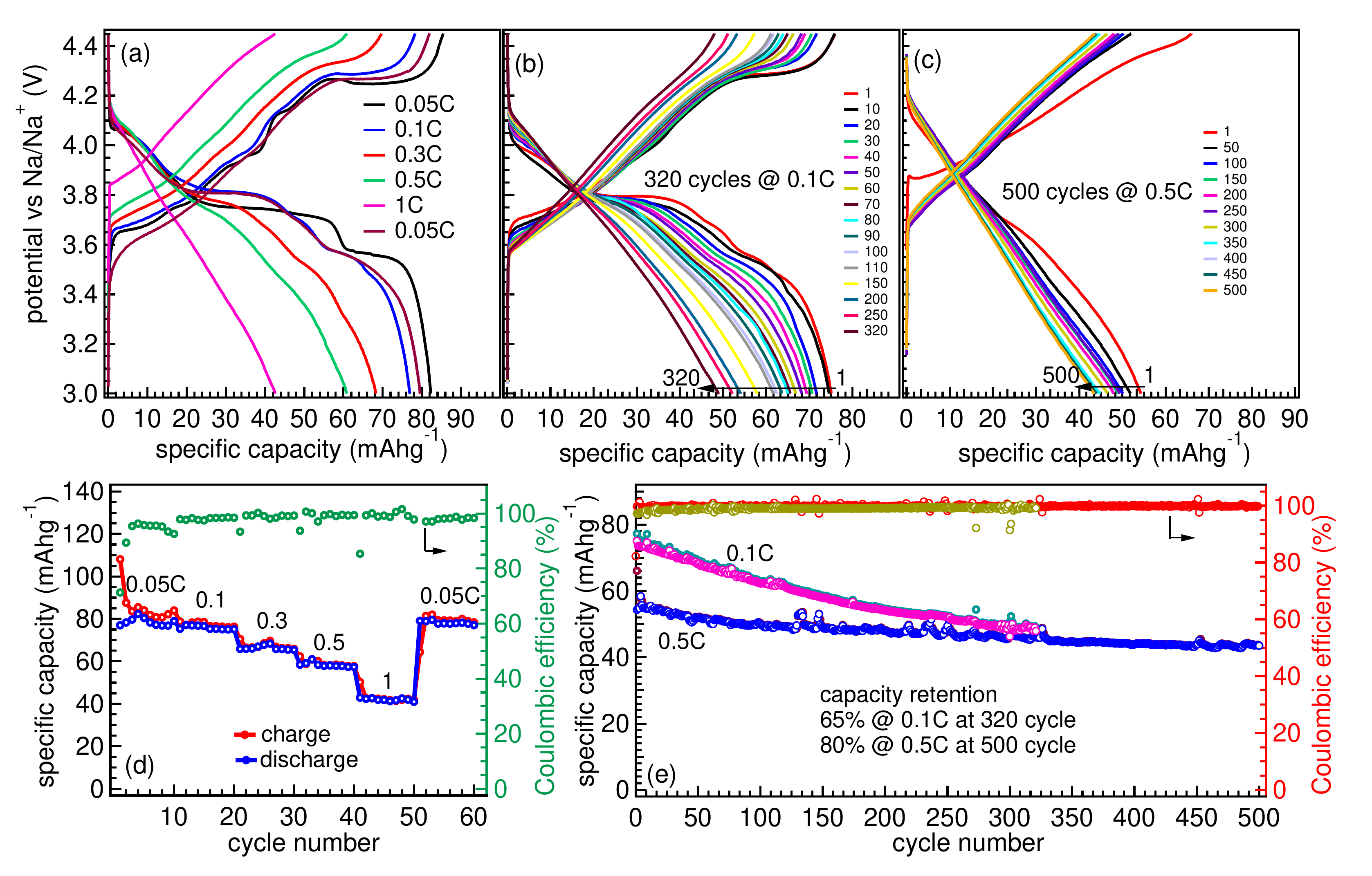}
\caption {The galvanostatic charge--dicharge profiles of Na$_{2}$Ni$_{2}$TeO$_{6}$ cathode in a voltage window of 3--4.45~V (a) at different current rates, (b) at 0.1~C current rate upto 320 cycles, (c) at 0.5~C current rate upto 500 cycles. (d) The rate capability behavior along the Coulambic efficiency for total 60 cycles at different current rates. (e) The specific capacity and Coulambic efficiency for the long cycling performance test at 0.1~C and 0.5~C rates upto 321 and 500 cycles, respectively.} 
\label{DP_O_K}
\end{figure*}

To differentiate the double phase and single-phase reactions in Na$_{2}$Ni$_{2}$TeO$_{6}$ cathode \cite{GrundishCM20}, we plot in Fig.~2(d) the voltage-composition curves at the scan rates of 0.05~C and 0.1~C. According to the profile at 0.05~C, extraction of 0.67 of Na$^{+}$ ions occur with a sequential phase transitions at 3.64, 3.76, 3.91, 3.95 and 4.13~V, generating an ordered phase of Na$_{1.33}$Ni$_{2}$TeO$_{6}$. This process refrains the mixing of transition metal ions with the vacant Na$^{+}$ site and facilitates the oxidation of 1/3 Ni$^{2+}$ to Ni$^{3+}$ per formula unit ($Z=$ 2). It was shown that the main contribution in electrochemical characteristics comes from the redox of Ni only as Te does not take part in the reaction. The inactiveness of Te in the electrochemical process is due to the fully filled $d-$orbitals of Te$^{6+}$, as a result the hybridization occurs mainly between oxygen and the Ni atoms \cite{GuptaJPS13, WangAM19}. Therefore, the double phase Na$_{2}$Ni$_{2}$TeO$_{6}$ - Na$_{1.33}$Ni$_{2}$TeO$_{6}$ attributes Na$^{+}$/vacancy ordering during charging from 3.5~V to 4.13~V. The plateau at 4.25~V accompanies 0.8 of Na$^{+}$ extraction forming a final product of Na$_{1.2}$Ni$_{2}$TeO$_{6}$ at this voltage. There is a de-insertion of 1.2 of Na$^{+}$ ion concentration at the end of charge (4.45~V) forming Na$_{0.8}$Ni$_{2}$TeO$_{6}$, which contains 0.8 Ni$^{2+}$ and 1.2 Ni$^{3+}$. This phase may involve a solid-solution phase regime owing to the formation of SEI layer. This regime does not support the formation of any ordered phase like Na$_{1.33}$Ni$_{2}$TeO$_{6}$ \cite{GuptaJPS13, WangNC15}. The discharge curve of Na$_{2}$Ni$_{2}$TeO$_{6}$ exhibits 4 plateaus at 3.54, 3.64, 3.8 and 4.07~V. An initial Na$^{+}$ insertion shows a sloppy feature (s-shaped) in the voltage window of 4~V to 3.64~V, indicating the formation of solid-solution regime, which may be stemmed due to the disordered phase of Na$_{2}$Ni$_{2}$TeO$_{6}$ cathode material upon insertion of Na$^{+}$. This trend also matches with CV profile, which shows relatively broad peaks at 3.94 and 3.84~V causing due to solid-solution state. There is a phase transition at 3.64~V involving the insertion of 0.42 Na$^{+}$ in the host structure. This feature attributes a quite sharp redox peak at 3.66~V than other cathodic peaks in CV. The repeated profile with similar redox plateaus/peaks in charge-discharge and CV characteristics show the stability of this cathode material for sodium-ion batteries \cite{GuptaJPS13, WangNC15}. The multi-plateau voltage profile can be supported by cyclic voltammetry profile recorded at scan rate of 0.05 mV s$^{-1}$, defining high reversibility of Na$_{2}$Ni$_{2}$TeO$_{6}$ cathode material. 

In Fig.~3(a) we show the galvanostatic charge-discharge profiles at different current rates, exhibiting maximum capacity at the corresponding C-rates. These results demonstrate the average mid-working potential variation with different C-rates from 3.78--3.93~V, which classify Na$_{2}$Ni$_{2}$TeO$_{6}$ as a high-voltage cathode material for SIBs. The rate capability at different current rates is presented in Fig.~3(d). The capacity is decreasing with increasing current rates (from 0.05 to 1~C) due to rate diffusion phenomenon of the material. However, we observe around 40~mAh g$^{-1}$ capacity and 3.93~V mid-working potential even at 1~C current rate. The high working potential is due to the inductive effect of TeO$_6$ octahedra owing to strong covalent nature of Te--O bond, which indicate the applicability of Na$_{2}$Ni$_{2}$TeO$_{6}$ material in high power applications \cite{GuptaJPS13, YangJPS17}. Finally to check the reversibility of electrode, we test again at 0.05~C rate, which shows the capacity similar to the initial value and confirm the high reversibility of this cathode material. 
In the rate capability test, we observe nearly 100\% Coulombic efficiency throughout depicting its stability at high current rates. Moreover, to check the capacity retention with long cycling we measured the GCD characteristics at 0.1~C and 0.5~C as shown in Figs.~3(b) and (c), respectively. Interestingly, we found the material highly stable with a significant capacity retention of 80\% after 500 cycles at 0.5~C and about 65\% after 320 cycles at 0.1~C maintaining nearly 100\% Coulombic efficiency in both the cases, as presented in Fig.~3(e). The data presented in Figs.~3(c, e) are performed on the same cell which was already tested for 60 cycles at different C values in Fig.~3(d). The stability here is found to be better than Na$_4$NiTeO$_6$ electrode, which shows 80\% retention in 100 cycles at the same current rate \cite{YangJPS17}. In fact, there is significant capacity fading within 100 cycles for Na$_3$Ni$_{1.5}$TeO$_6$ electrode when cycled at low current rate \cite{GrundishCM20}.  

 \begin{SCfigure*}
\includegraphics[width=5.75in]{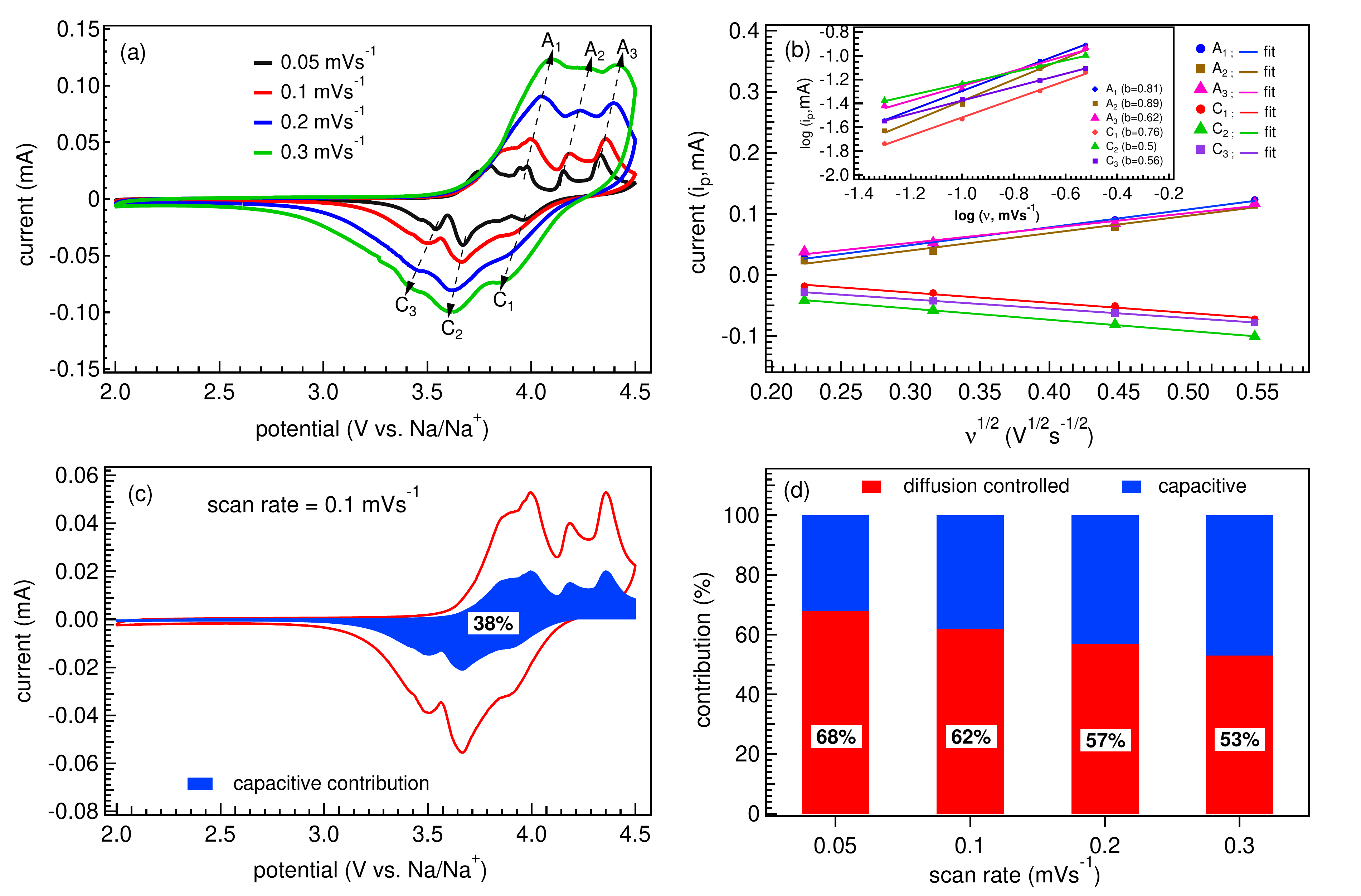}
\caption {(a) The CV curves of Na$_{2}$Ni$_{2}$TeO$_{6}$ at different scan rates of 0.05, 0.1, 0.2 and 0.3 mV s$^{-1}$ having three anodic peaks (marked by A$_{1}$, A$_{2}$, A$_{3}$) and three cathodic peaks (marked C$_{1}$, C$_{2}$, C$_{3}$. (b) The linear fit between peak current i$_{p}$ and square root of scan rate for different scan rates. (c) The shaded area showing capacitive contribution at a scan rate of 0.1 mV s$^{-1}$, and (d) the capacitive and diffusive contributions for the total current at different scan rates.}
\label{DP_O_K}
\end{SCfigure*}

\subsection{\noindent ~Analysis of cyclic voltammetry:}

The reduction/oxidation characteristics, phase transformations and diffusion coefficient of Na$_{2}$Ni$_{2}$TeO$_{6}$ electrode reaction have been investigated by the CV measurements against Na/Na$^{+}$ in half-cell configuration. Fig.~4(a) shows the CV curves (3rd cycle) at different scan rates from 0.05--0.3 mV s$^{-1}$ in a voltage range of 2--4.5~V. We can clearly observe the disappearance of some initial anodic peaks at higher scan rates of 0.2 and 0.3 mV s$^{-1}$, which indicates absence of some redox reactions due to lack of time for their completion. Furthermore, the anodic peak gradually shifts to higher potential and the respective cathodic peaks move towards lower potential as the scan rate increases, depicting some irreversibility features of cathode material due to the faster rate kinetics of the species \cite{YuanJMCA13}. For further analysis, in Fig.~4(b) we plot the peak current i$_{p}$ versus square root of scan rate ($\nu$$^\frac{1}{2}$) for three anodic and cathodic couples, marked by (A$_{1}$, A$_{2}$, A$_{3}$) and (C$_{1}$, C$_{2}$, C$_{3}$), respectively. In order to calculate the sodium ion diffusion coefficient values, the relationship between $i_{p}$ vs. $\nu$$^\frac{1}{2}$ is described by the Randles-Sevcik equation for the diffusion controlled electrochemical reaction \cite{XiaoAEM19}, as described below:
 \begin{eqnarray}
 i_{p} = (2.69 \times 10^{5}) A D^{\frac{1}{2}}  C n^{\frac{3}{2}} \nu^{\frac{1}{2}},
 \end{eqnarray}
here $i_{p}$ is the peak current (mA), A is the area of the electrode (1.13 cm$^{2}$), C is the bulk concentration of the Na ions in the electrode (0.0014 mol cm$^{-3}$), $n$ is the no of the electrons transferred in an electrochemical reaction (1.2), $\nu$ is the scan rate (mV s$^{-1}$) and $D$ is the diffusion coefficient of the sodium ions in the cathode (cm$^{2}$ s$^{-1}$). The slope of a linear fit to the plots of $i_{p}$ vs. $\nu$$^\frac{1}{2}$ has been used to calculate the $D$ values for different peaks. The estimated values of $D$ for different cathodic and anodic peaks varies in the range of 2$\times$10$^{-10}$ to 5$\times$10$^{-12}$ cm$^{2}$s$^{-1}$. 

Further, using the CV at different scan rates, the difference between the diffusion and surface-controlled reactions can be obtained with the assumption that current obeys power law relationship as written below \cite{LiuAS18}:
\begin{eqnarray}
i = a\nu^{b}
\end{eqnarray}
here $i$ is the current, $\nu$ is the scan rate, $a$ and $b$ are the parameters, where $b$ is calculated from the slope of the linear plot of log $i$ vs log $\nu$, as shown in the inset of Fig.~4(b) for the anodic and cathodic peaks. The value of $b$ is crucial here, as $b=$ 0.5 corresponds to the diffusion controlled reaction and $b=$ 1 corresponds to the capacitive (surface controlled) reaction mechanism, whereas the $b$ values in between 0.5 and 1, imply for the pseudo-capacitive behavior for the sodium ion storage \cite{ValvoEA18}. The obtained $b$ values for A$_{1}$, A$_{2}$, A$_{3}$, C$_{1}$, C$_{2}$ and C$_{3}$ peaks are as 0.81, 0.89, 0.62, 0.76, 0.5 and 0.56, respectively. These results affirm the mixed behavior of  Na$_{2}$Ni$_{2}$TeO$_{6}$, i.e. both diffusion controlled and pseudo-capacitive mechanisms exist in the storage process. Notably, this type of mixed behavior provides efficient Na$^{+}$ storage in the host structure, resulting high capacity retention during long cycling. 

\begin{figure*}
\includegraphics[width=6.8in]{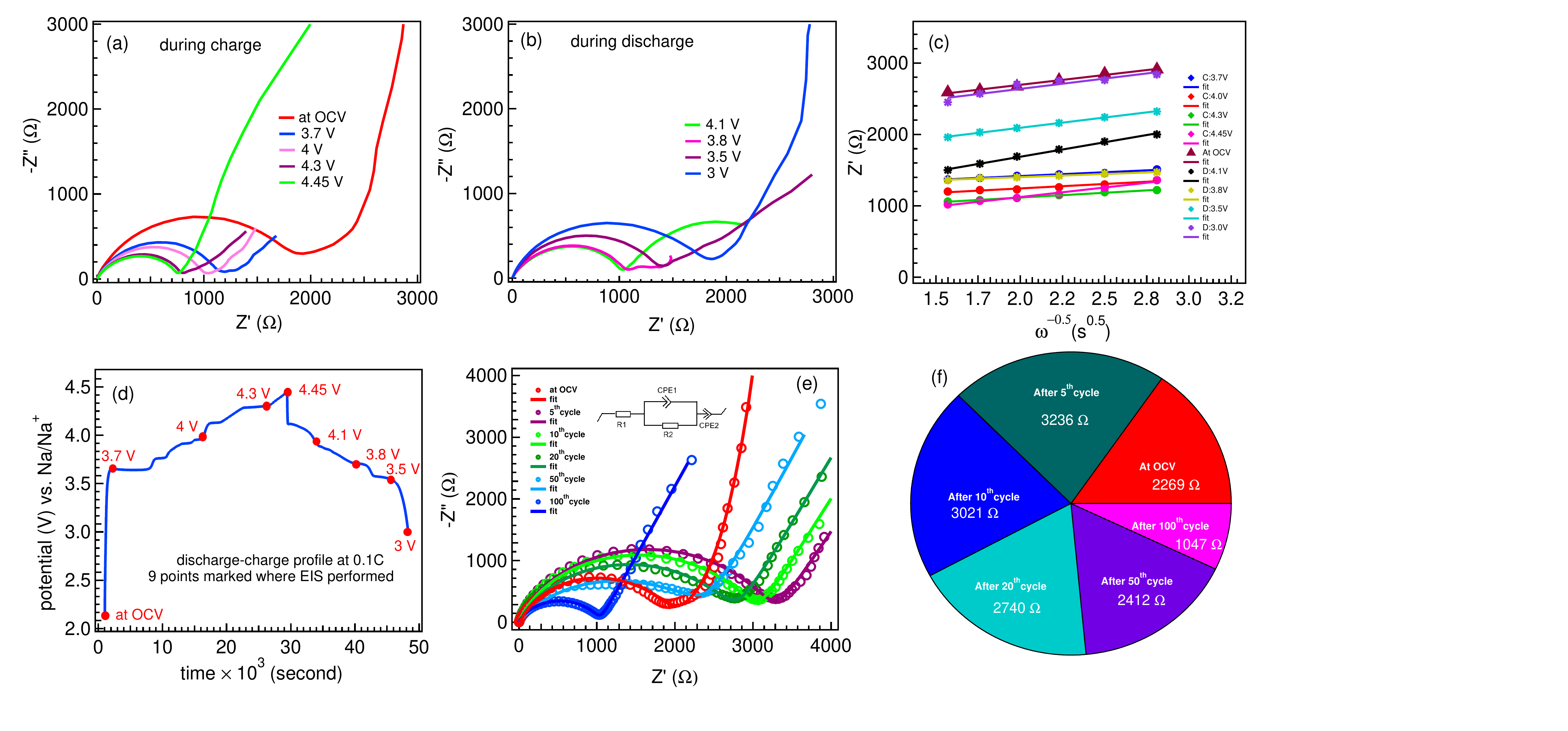}
\caption {The EIS spectra of the Na$_{2}$Ni$_{2}$TeO$_{6}$ cathode material collected at (a) different charging states from OCV to 4.45~V (b) different discharging states from 4.1 to 3~V. (c) The linear fit between $\omega$$^{-0.5}$ vs Z$_{re}$ at different charged states to determine the slope $\sigma$ to extract the diffusion coefficient. (d) The galvanostatic profile of charging-discharging with points indicated where impedance measurements were done, and (e) the equivalent circuit fitted Nyquist plots after 5, 10, 20, 50 and 100 cycles. (f) The variation of impedance parameter, $R_{\rm ct}$ values extracted by fitting the EIS data in (e) during cycling.} 
\label{DP_O_K}
\end{figure*}

It is important to quantify the diffusion controlled and capacitive controlled charge storage contributions at a fixed voltage. In this regard the response current can be expressed as a contributions from both the capacitive and diffusive process of the material, as given below \cite{XieAM17, ChaoNC16}:
 \begin{eqnarray}
i({\rm V}) = k_{1} \nu+k_{2} \nu^{1/2},
 \end{eqnarray}
here $k_1$$\nu$ and $k_2$$\nu$$^{1/2}$ correspond to the capacitive and diffusion-controlled contribution, respectively. The linear relation between $i$(V)/$\nu$$^{1/2}$ and $\nu$$^{1/2}$ gives the slope $k_1$ and intercept $k_2$ at each voltage. A capacitive contributed area of 38\% is shaded in the CV curve at 0.1 mV s$^{-1}$ scan rate in Fig.~4(c) for better representation of distribution pattern of total current. In Fig.~4(d) we show the extracted contribution of the capacitive and diffusion behavior in percentage ratio for the Na$_{2}$Ni$_{2}$TeO$_{6}$ electrode reaction at different scan rates. The diffusion current contributes 68\% of the total current at 0.05 mV s$^{-1}$ scan rate, whereas this contribution decreases gradually to 62\%, 57\% and 53\% in case of 0.1, 0.2 and 0.3 mV s$^{-1}$, respectively. This percentage of diffusion-controlled process follows a decreasing trend from lower to higher scan rates, which affirms the occurrence of more surface-controlled process at higher scan rates \cite{ZhangCEJ21}. These results suggest that the layered structure of Na$_{2}$Ni$_{2}$TeO$_{6}$ cathode material provides more active surface area and open diffusion path for sodium ions in the square shaped space between four oxide ions, which attributes for surface controlled reaction with pseudo-capacitive nature \cite{YabuuchiCR14}. 

\subsection{\noindent ~{\it In-situ} electrochemical impedance spectroscopy:}

Now we present the {\it in-situ} electrochemical impedance spectroscopy (EIS) measurements to determine the effect of charge-discharge voltage and cycling on the charge transfer resistance as well as on sodium-ion diffusion in Na$_{2}$Ni$_{2}$TeO$_{6}$ electrode. 
Figs.~5 (a, b) show the EIS curves recorded at open circuit voltage (OCV) and at different charge/discharge potentials, as marked in Fig.~ 5(d). The EIS measurements at different number of cycles and OCV are compared in Fig.~5(e) with the corresponding equivalent circuit in the inset. Here the impedance spectra include a semi-circle in the high-mid frequency zone representing the charge transfer resistance ($R_{\rm ct}$) and a low frequency step inclined line to depict the bulk diffusion of Na-ions in the Na$_{2}$Ni$_{2}$TeO$_{6}$ cathode material. The overall impedance of Na$_{2}$Ni$_{2}$TeO$_{6}$ cathode vs. Na$^{+}$/Na decreases with the increase in charging potential, and impedance again increases with decreasing potential during discharging, as clearly visible in Figs.~5(a, b). In case of charging, the decreasing trend of $R_{\rm ct}$ at higher voltages implies more reactions between electrode and electrolyte and providing greater surface stability during de-sodiation \cite{ChenJMCA15}. Owing to this stable structure, this cathode exhibits significant capacity retention, as shown in Fig.~3, by reversible accommodation of Na$^{+}$ ions during discharge. Furthermore, the dominance of electron transfer reaction over the de-sodiation of Na$^{+}$ may decrease the $R_{\rm ct}$ value at higher charging potentials, i.e., at 3.8~V and 4.45~V \cite{RamasamyPCL17}. This phenomenon results due to the oxidative decomposition of as-formed SEI layer, which gradually conducts electron during de-sodiation at higher voltages. In Fig.~5(b) the gradual increase in impedance value ($R_{\rm ct}$) during discharging from 4.1 to 3~V is clearly visible. The de-sodiation process involves formation of insulating and thicker SEI layer associated with progressive volume change. Therefore, the $R_{\rm ct}$ increases at lower  potentials during discharge due to the blocking effect of SEI layer, which results in high kinetic barrier for the infiltration of Na$^{+}$ ions through the SEI layer. In the discharge process, there is a gradual movement of the sloppy line in the low frequency region towards the real axis and at 3.5~V it becomes straight resulting the low frequency tail closer to real axis. This type of behavior clearly indicates the violation of semi-infinite diffusion theory due to the sluggish mass diffusion condition. 

 \begin{SCfigure*}
\includegraphics[width=5.73in]{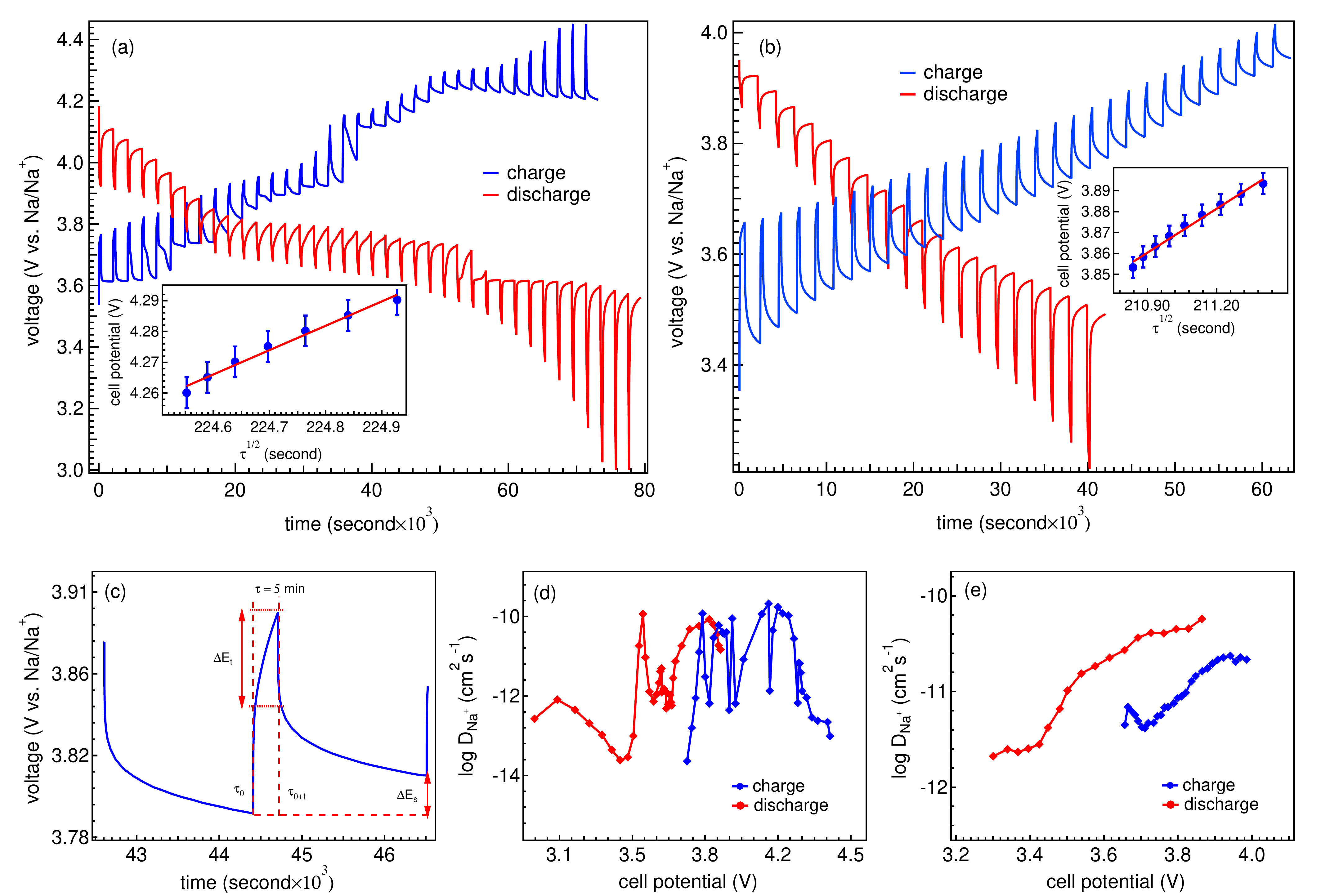}
\caption {The GITT profiles of Na$_{2}$Ni$_{2}$TeO$_{6}$ electrode at a current density of 0.1~C (a) during 2nd cycle in a voltage window of 3--4.45~V, (b) during 502th cycle in a voltage window of 3.21--4.01~V. (c) A single titration curve at 3.89~V during charging condition, showing excitement and relaxation process. The Na-ion diffusion coefficients as a function of cell potential deduced by the GITT method during (d) 2nd cycle (e) 502th cycle. Insets in (a, b) show the linear relation between voltage and $\tau^{1/2}$.} 
\label{DP_O_K}
\end{SCfigure*} 

Moreoevr, to understand the diffusion kinetics at different charge states we extract the diffusion coefficient of Na$^+$ using the equation below \cite{BaiAMI16, LiCEJ17}:
\begin{eqnarray}
D=1 /2 [(R\times T)/AF^{2}n^{2}C\sigma]^{2}
 \end{eqnarray}
here the related parameters are $R$ (gas constant), $T$ (absolute temperature), $A$ (area of the electrode), $n$ (number of electrons), $F$ (Faraday's constant), and $C$ (concentration of Na$^{+}$), respectively. Also, the Warburg impedance factor and the Z$_{re}$ shows the relation given below:
\begin{eqnarray}
{Z}^\prime = R_{s} + R_{ct} + \sigma \omega^{-0.5}
 \end{eqnarray} 
 and here the R$_s$ and $\omega$ represent the ohmic resistance (or solution resistance) and angular frequency, respectively. The value of $\sigma$ can be obtained from the plot between Z$_{re}$ and the reciprocal square root of the angular frequency ($\omega$$^{-0.5}$) by using the slope at different charge and discharge states, as shown in Fig.~5(c). The calculated diffusion coefficient of Na$^{+}$ at OCV is found to be 0.54$\times$10$^{-12}$ cm$^{2}$ s$^{-1}$, which shows sluggish diffusion of Na$^{+}$ at OCV. The value of $D_{Na+}$ at different charge/discharge states falls in the range of 0.5--3.6$\times$ 10$^{-12}$ cm$^{2}$ s$^{-1}$. Furtehr, the EIS study in the frequency range 10 mHz to 100 KHz is compared at OCV and at different cycles, i.e., at 5, 10, 20, 50 and 100 cycles (at fully discharged state). The inset of Fig.~5(e) denotes the corresponding equivalent circuit, where R1 represents the solution resistance or ohmic resistance, the parallely connected constant phase element unit (CPE1) and the resistor (R2) attributes the charge transfer resistance ($R_{\rm ct}$) and the series connected CPE2 signifies the capacitive-controlled process \cite{WuAMI17}. At OCV, the $R_{\rm ct}$ is evaluated to be 2269 $\Omega$, which increases to 3236~$ \Omega$ after 5th cycle. This increment of R$_{ct}$ value suggests the continuous growth of SEI layer from OCV up to 5th cycle \cite{SongAEM18}. After that the R$_{ct}$ value decreases to 3021, 2740, 2412 and 1047~$\Omega$ at 10, 20, 50 and 100 cycles, respectively, as shown in Fig.~5(f). This reduction in R$_{ct}$ may be due to the formation of stable surface film leading to lower over-potential during cycling process \cite{SongAEM18}. 

\subsection{\noindent ~Galvanostatic intermittent titration technique:}

Note that the predominant process of Na-intercalation/de-intercalation during cycling involves Na$^{+}$ diffusion coefficient, which is a critical dynamic parameter to measure the migration rate of Na$^{+}$ ions into the host material \cite{WangNano18}. In order to get more insight of diffusion kinetics we use a chronopotentiometry based technique, namely GITT under the thermodynamic equilibrium conditions. In the GITT measurements, a constant current is applied to the electrode for short duration, then the electrode is allowed to relax to a steady state value after reaching OCV \cite{MaheshEA20}. Herein, the cells were first charged to maximum cut-off voltage with constant current density of 0.1~C for a duration of 5 min, which was followed by an OCV stand of 30 min to reach a steady state value ($E_{\rm s}$), then the cells were discharged with the same condition up to the minimum discharge voltage. Figs.~6(a, b) depict the GITT curves of Na$_{2}$Ni$_{2}$TeO$_{6}$, during the 2nd cycle for a cut-off voltage window of 3--4.45~V and during 502th cycle in a voltage window of 3.21--4.01~V, respectively. The curves in Fig.~6(a) follows the multi-plateau behavior as in galvanostatic charge-discharge profile at 0.1~C, and gradually disappears in the GITT curves measured during 502th cycle [see Fig.~6(b)]. The magnification of single titration curve is represented in Fig.~6(c), which shows the change in the cell voltage during the 5 min current pulse from $\tau$$_{0}$  to $\tau$$_{0+t}$ ($\Delta$ $E_{\tau}$) and the variation of cell potential during the relaxation period of 30 min ($\Delta$ $E_{\rm s}$). The variation of voltages at each step are evaluated as a function of time and the diffusion coefficient of Na$^{+}$ in the host material can be calculated from the following equation, which relies on Fick’s second law \cite{WangACS17}.
\begin{eqnarray}
	D_{Na^{+}}= \frac{4}{\pi \tau}\left[\frac{m_{\rm B} V_{\rm M}}{M_B A}\right]^2\left[\frac{\Delta E_s}{\tau(\Delta E_t/d\sqrt \tau)}\right]^2
	\end{eqnarray}                                                                       
Here, m$_{B}$(g) and $M_{\rm B}$(387 g mol$^{-1}$) are the mass and molecular weight of the active materials in the electrode, $V_{\rm M}$(78.6 cm$^{3}$ mol$^{-1}$ per formula unit) and $A$ (cm$^{2}$) are the molar volume and the total area of the electrode, $d$ is the diffusion length and $\tau$ is the duration of current pulse (5 min). When the steady state voltage varies linearly with $\tau$$^{1/2}$, as shown in the insets of Figs.~6(a, b), then the eq.~6 can be expressed in simple form as  below \cite{ZhengESM18}: 
\begin{eqnarray}
	D_{Na^{+}}= \frac{4}{\pi \tau}\left[\frac{m_B V_M}{M_B A}\right]^2\left[\frac{\Delta E_s}{\Delta E_t}\right]^2
	\end{eqnarray}                                                                                                                 
The above equation considers a stable molar volume during sodiation and de-sodiation in the electrode material. The resulting Na$^{+}$ diffusion coefficients for 2$^{nd}$ cycle and 502$^{th}$ cycle are presented in Figs.~6(d, e) as Log D$_{Na}$$^{+}$ versus cell potential for both charging and discharging states. The growth profiles of diffusion coefficient show more minima points for 2$^{nd}$ cycle as compared to 502$^{th}$ cycle, which suggests sluggish diffusion kinetics in case of the former. The minima points in Log D$_{Na}$$^{+}$ versus potential curves indicate phase transitions owing to strong interactions between intercalated Na$^{+}$ ions and Na$_{2}$Ni$_{2}$TeO$_{6}$ host species or degree of disorderness of host species during cycling \cite{BucherCM16}. The diffusion coefficient values extracted from GITT measurements and using equation 7 lies in the range of 10$^{-10}$ to 10$^{-12}$ cm$^{2}$s$^{-1}$ at different potentials during 502 cycle. 

Intriguingly, the values of diffusion coefficient extracted by CV, EIS and GITT analysis are consistent, and indicate relatively fast mobility of sodium ions in Na$_2$Ni$_2$TeO$_6$, which found to be comparable with reported in \cite{WangNano18} for Na$_3$Ni$_2$SbO$_6$ and in \cite{SongJMCA14} for very famous Na$_3$V$_2$(PO$_4$)$_3$ polyanionic electrode. Further, the diffusion kinetics support for the stability and high-capacity retention of the material during long cycling. Note that the ion transport pathways were found to be highly anisotropic by molecular dynamics simulation \cite{SauJPCC15_NNTO, SauJPCC15_NMTO}. In fact, Bera {\it et al.} performed detailed neutron powder diffraction measurements and used bond valence sum analysis to find the  accessible sites and pathways for sodium-ion conduction in Na$_2$Ni$_2$TeO$_6$ \cite{BeraJPCC20}. The conduction of Na ions was observed only through two-dimensional (2D) pathways within the Na layers in the $ab$ plane. However, the edge sharing (Te/Ni)O$_6$ octahedra are so closely packed, so it is impossible for Na ions to move through these intervening metal--oxide layers along the $c$-direction. There are three different crystallographic sites for sodium occupation and the disorderness in Na-ion distribution provides such type of high ionic conductivity \cite{SauJPCC15_NNTO, SauJPCC15_NMTO}. Further, it was found that the conductions of sodium-ion within the 2D $ab$ plane are through Na1 and Na2 sites in zigzag-like pathways \cite{BeraJPCC20}. As no other ions present within the Na layers, there are no collisions/scattering in the pathways of conduction through $ab$ plane, which leads to the higher ionic conductivity. Our results demonstrate high cyclic/structural stability of this electrode even better at higher current rate, which plays a vital role for its practical applications in batteries to power portable devices for long time as well as for fast charging of the devices. 

\subsection{\noindent ~During and post cycle analysis:}

Finally, in order to understand the post cycling changes in the structural and morphology, we dismantle the cells after 10 cycles at 0.1~C and after 100 cycles at 0.5~C. The recovered electrode materials were washed with diethyl carbonate (DEC) twice for removing unreacted electrolyte, and dried in argon gas filled glove box for 48 h. 
 \begin{figure}[h]
\includegraphics[width=3.4in]{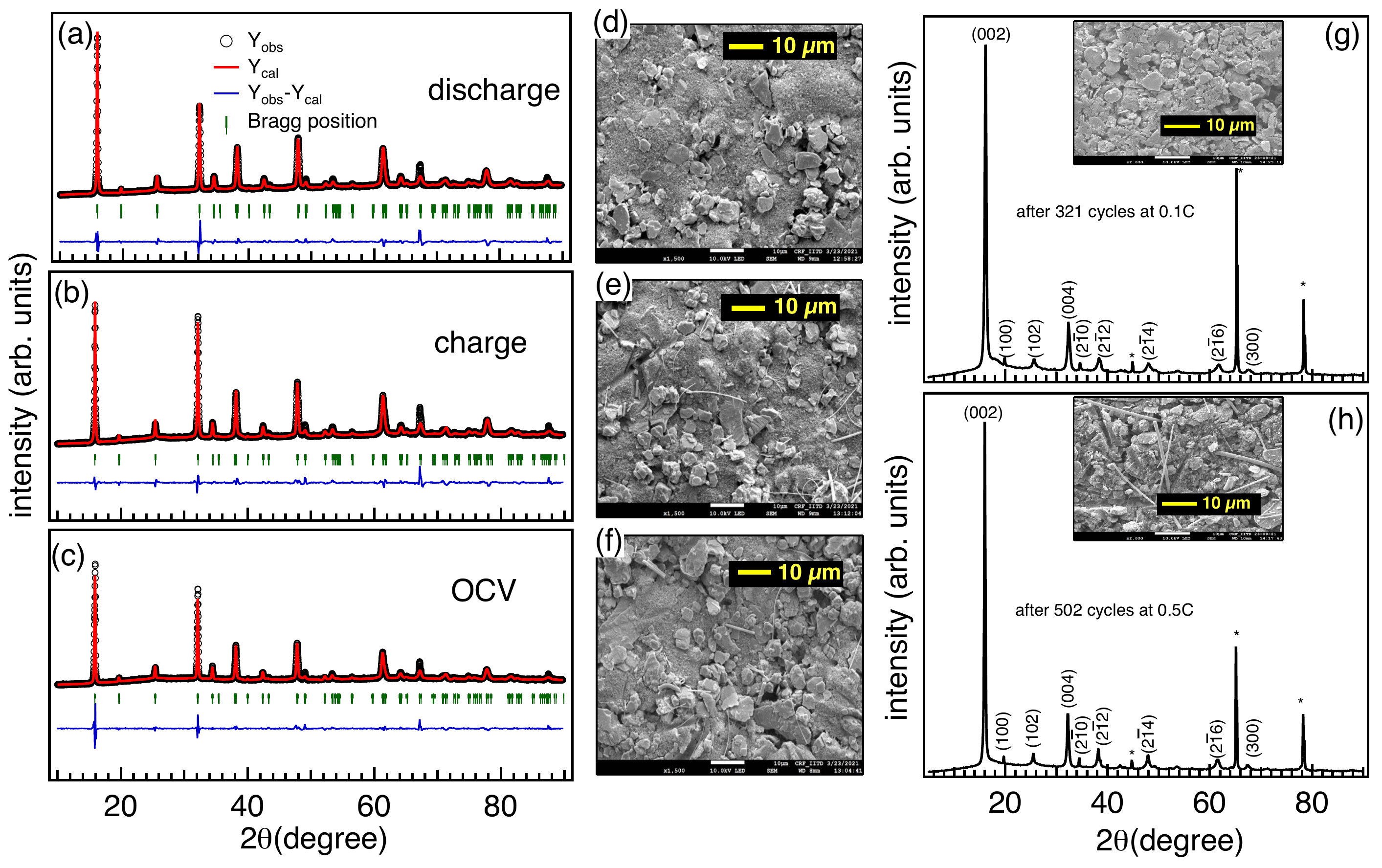}
\caption {Rietveld refinement of {\it ex-situ} XRD patterns collected from the cells at fully discharged (a), fully charged (b) and at OCV (c) states. The morphology of (d) fresh electrode, (e) after 10 cycles at 0.1~C, and (f) after 100 cycles at 0.5~C. The XRD patterns measured (g) after 321 cycles at 0.1~C and (h) after 502 cycles at 0.5~C, the peaks marked by * are from Al.} 
\label{DP_O_K}
\end{figure} 
In Fig.~7, we present the morphologies of fresh electrode before electrochemical analysis (a), after 10 cycles at 0.1~C (b) and after 100 cycles at 0.5~C (c), which found to be similar without any cracks and discontinuity. The results show high stability of the electrode with electrolyte which can be inferred for long cycle life of the battery as shown by the capacity retention in Fig.~3. However, some fibres can be seen in the images of cycled electrodes stuck from separator. For the post-mortem structural analysis of the electrode using {\it ex-situ} XRD, we retrieve the materials at open circuit voltage (OCV), fully charge stage and fully discharge stage, as shown in Figs.~7(a--c), respectively. The structural parameters extracted from the Rietveld refinement of these XRD patterns are found to be during charge ($a=b=$ 5.201~\AA~and $c=$11.131~\AA), discharge ($a=b=$ 5.199~\AA~and $c=$11.123~\AA)~and at OCV $a=b=$ 5.205~\AA~and $c=$11.139~\AA). The XRD patterns recorded in all three stages look similar without significant change in the lattice parameters, supporting structural stability of Na$_{2}$Ni$_{2}$TeO$_{6}$ cathode in charge/discharge stages. Moreover, we perform the {\it ex-situ} XRD measurements of extracted cathode material after 321 cycles at 0.1~C and 500 cycles at 0.5~C, as shown in Figs.~7(g, h), respectively. There are no significant changes in the XRD patterns except the change in the peak intensity, which indicate the stable and reversible structural geometry of this cathode after long cycling in Na-half cell configuration.

Further, the electronic properties of Na$_{2}$Ni$_{2}$TeO$_{6}$ electrode were investigated using x-ray photoelectron spectroscopy (XPS). In Fig.~8, we show the Ni 2$p$, Te 3$d$, Na 1$s$ and O 1$s$ core-level spectra of pristine sample (a1--a4), fully discharged (b1--b4) and fully charged (c1--c4) states. The peak positions are calibrated by consider the binding energy of C 1s core-level at 284.8~eV. We use Voigt function having both the Lorentzian and Gaussian broadenings for deconvolution of the core-level peaks with Tougaard background subtraction \cite{Dhaka_PRB_11}. For the pristine sample, the fitted Ni 2$p$ spectrum in Fig.~8(a1) consists of two spin-orbit components at 855~eV and 873.2~eV corresponds to 2p$_{3/2}$ and 2p$_{1/2}$,  respectively, which confirm the presence of Ni$^{2+}$ state. In addition, there are two shake up satellite peaks observed at higher binding energies of 861.3~eV and 879.1~eV, which are consistent with reported in refs.~\cite{YangJPS17}. Interestingly, in case of the fully charged state the deconvolution of the Ni 2$p$ core-level reveals the appearance of nearly 60\% Ni$^{3+}$ valence state. This dominant contribution of Ni$^{3+}$ is in agreement with the electrochemical results presented in Fig.~2(d), where total 1.2 Na$^{+}$ are de-intercalated from the structure during charging. In the fully discharged state, the spectra contain only Ni$^{2+}$ state elucidating the conversion of Ni$^{3+}$ to Ni$^{2+}$ due to sodium insertion in the material, which is expected to be the case \cite{GuptaJPS13}. For the Te 3$d$ core-levels, the peaks at 576~eV and 586.4~eV correspond to 3d$_{5/2}$ and 3d$_{3/2}$, respectively, representing the presence of Te$^{6+}$ state \cite{YangJPS17}. Moreover, the Na 1$s$ spectrum ensures the Na$^{+}$ state, situated at 1071~eV binding energy. Both the Te 3$d$ and Na 1$s$ spectra show no shift during charging-discharging process, affirming the inactiveness of these elements in redox reactions. In case of O 1$s$, the core-level spectra attribute three peaks at 530, 531.5 and 534.5~eV for the pristine sample, where the first peak is ascribed to the O$^{2-}$ anion in the crystalline domain and the other peaks determine the weakly adsorbed surface species i.e. hydroxylation and carbonation of the surface species \cite{AnNE14}. For the charged state, there is an evolution of another peak at 530.8~eV contributing the O$^{1-}$ species, which is originated due to the degradation of lattice oxygen O$^{2-}$ from the bulk surface \cite{DongJMCA15}. Notably, the disappearance of O$^{1-}$ peak during the course of discharge, assures the participation of O$^{1-}$/O$^{2-}$ redox couple in Na-ion insertion/de-insertion process \cite{YabuuchiJACS11}. However, the increment in intensity and shift in the position of the 534.5~eV peak signifies gradual degradation of electrolyte during charging/discharging. Also, the Na KL$_1$L$_{23}$ auger peak is observed at $\approx$535.5~eV for Al K$\alpha$ x-ray source \cite{DubeyACS21}. 

 \begin{figure}
\includegraphics[width=3.5in]{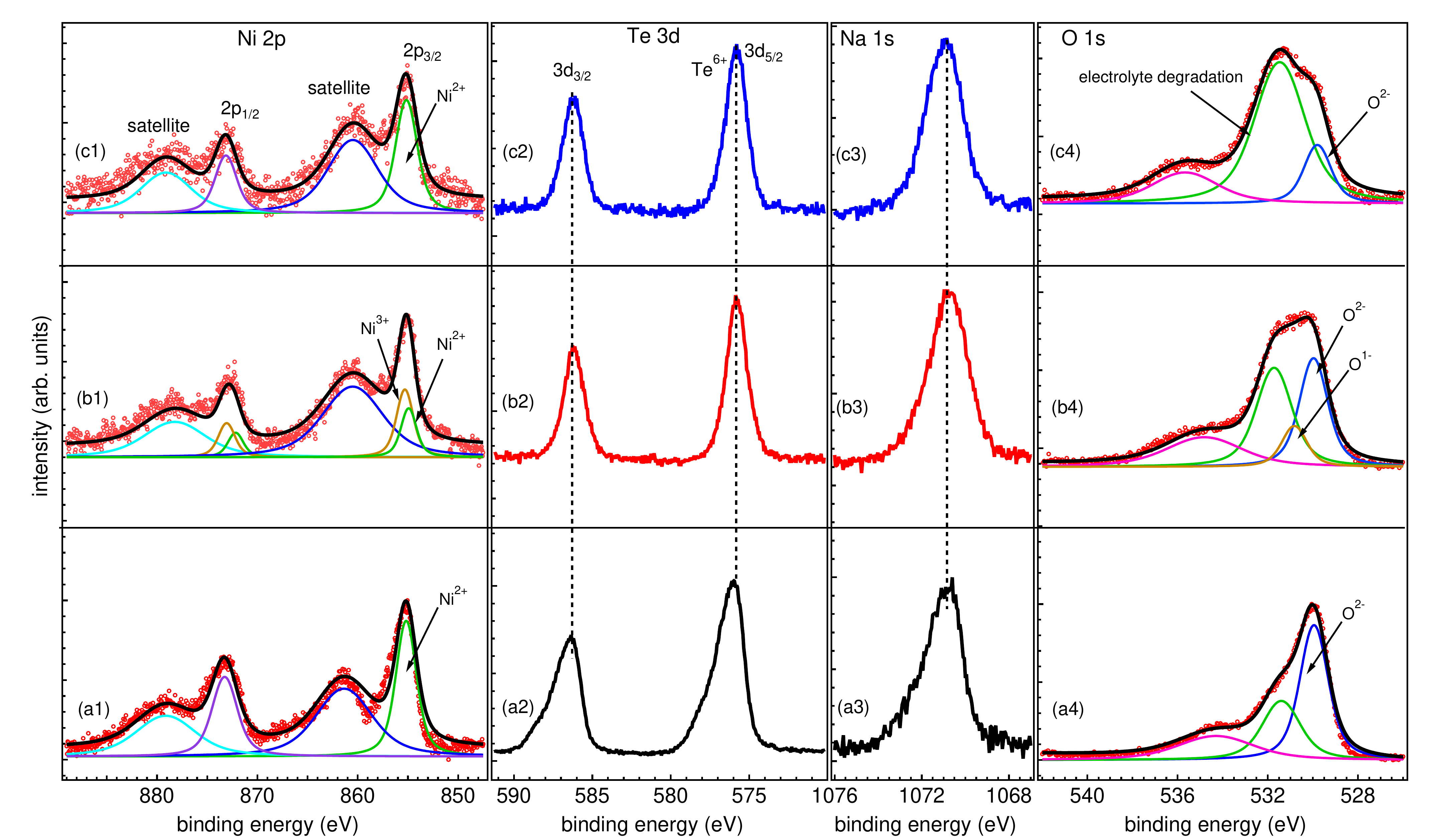}
\caption {The Ni 2$p$, Te 3$d$, Na 1$s$ and O 1$s$ XPS core-level spectra of Na$_2$Ni$_2$TeO$_6$ electrode material, (a1--a4) for pristine, (b1--b4) fully discharged, and (c1--c4) fully charged states.} 
\label{DP_O_K}
\end{figure}

\section{\noindent ~Conclusion}

In conclusion, we report electrochemical study and diffusion kinetics of honeycomb structured Na$_{2}$Ni$_{2}$TeO$_{6}$ as high voltage and stable cathode material for sodium-ion batteries. The XRD patterns have confirmed pure phase of the sample and the structural parameters extracted by Rietveld refinement are consistent with HR-TEM and SEAD studies. Interestingly, the discharge capacities of 82 and 77 mAhg$^{-1}$ are observed 0.05~C and 0.1~C current rates, respectively. The cathode shows excellent rate capability performance tested at various current rates along with mid-working potential of $\approx$3.9~V at 1~C. The capacity retention was found to be 
80\% after 500 cycles having nearly 100\% Coulombic efficiency at 0.5~C, which demonstrate the long cyclic stability of this cathode in sodium ion batteries useful for powering portable devices. The de-insertion/insertion of Na$^+$-ions during electrochemical cycling is consistent with the observed ratio of Ni$^{3+}$/Ni$^{2+}$ valence state in photoemission study. The detailed analysis of cyclic voltammetry at different scan rates suggest pseudo-capacitive nature of the cathode for sodium-ion storage. Moreover, the {\it in-situ} EIS studies reveal the behavior of charge-transfer resistance of this cathode at different charge/discharge stages as well as with different number of cycles. We have determined the diffusion coefficient to understand the kinetics using CV, EIS and GITT analysis, which found mainly in the range of 10$^{-10}$ to 10$^{-12}$ cm$^{2}$ s$^{-1}$ depending on the experimental parameters. Notably, the post-mortem analysis of electrodes using {\it ex-situ} XRD and FE-SEM establish its high structural and morphological stability after various charge-discharge cycles. Our detailed analysis demonstrates the capabilities of Na$_{2}$Ni$_{2}$TeO$_{6}$ cathode in sodium-ion batteries to power portable devices with long cyclic life.

%\section*{\noindent ~Author contribution statement}

%All the authors have contributed to this work.

%\section*{\noindent ~Declaration of Competing Interest}

%The authors declare that they have no known competing financial interests or personal relationships that could have appeared to influence the work reported in this paper.

\section{\noindent ~Acknowledgments}

Authors at IIT Delhi acknowledge the financial support from DST through ``DST--IIT Delhi Energy Storage Platform on Batteries" (project no. DST/TMD/MECSP/2K17/07) and from SERB-DST through core research grant (file no.: CRG/2020/003436) to set-up the facilities for sodium-ion battery project. J.P., H.R., and S.K.S. thank UGC, DST (project: DST/TMD/MECSP/2K17/07), and MHRD, respectively for the fellowship. We thank IIT Delhi for providing research facilities for sample characterization (XRD and Raman at the physics department, and FE-SEM \& HR-TEM at CRF).

\end{document}